\documentclass[useAMS,usenatbib]{mn2e}
\usepackage{amsmath,fleqn,graphicx,amssymb,txfonts}

\newcommand{\B}[1]      {\mbox{\boldmath$#1$}}
\newcommand{\bx}        {\B{x}}
\newcommand{\bv}        {\B{v}}
\newcommand{\ba}        {\B{a}}
\title{Initial conditions for disc galaxies}

\author[P.~J.~McMillan \& W.~Dehnen]{
  Paul~J.~McMillan$^{1,2,3}$\thanks{E-mail:p.mcmillan1@physics.ox.ac.uk}
  and  Walter~Dehnen$^{1}$\\
  $^{1}$Department of Physics \& Astronomy, 
  University of Leicester, Leicester, LE1 7RH, UK\\
  $^{2}$LAM, Observatoire Astronomique de Marseille Provence,
  2 Place Le Verrier,
  F-13248 Marseille Cedex 4, France\\ 
  $^{3}$Rudolf Peierls Centre for Theoretical Physics, 1 Keble Road,
  Oxford, OX1 3NP, UK
}

\begin{document}

\maketitle

\begin{abstract}
  We present a general recipe for constructing $N$-body realizations
  of galaxies comprised of near-spherical and disc components.  First,
  an exact spherical distribution function for the spheroids (halo \&
  bulge) is determined, such that it is in equilibrium with the
  gravitational monopole of the disc components. Second, an $N$-body
  realisation of this model is adapted to the full disc potential by
  growing the latter adiabatically from its monopole.  Finally, the
  disc is sampled with particles drawn from an appropriate
  distribution function, avoiding local-Maxwellian approximations. We
  performed test simulations and find that the halo and bulge radial
  density profile very closely match their target model, while they
  become slightly oblate due to the added disc gravity. Our findings
  suggest that vertical thickening of the initially thin disc is
  caused predominantly by spiral and bar instabilities, which also
  result in a radial re-distribution of matter, rather than scattering
  off interloping massive halo particles.
\end{abstract}

\begin{keywords}
  Methods: $N$-body simulations -- Galaxies: kinematics and dynamics
  -- Galaxies: haloes
\end{keywords}
\section{Introduction} \label{sec:ics:intro} Generating an equilibrium
$N$-body representation of a multi-component galaxy is of importance
for a number of applications, for example the study of bars
\citep[e.g.][]{DebattistaSellwood2000,Athanassoula2002}, warps
\citep[e.g.][]{Idetaetal2000} and galaxy mergers, both minor
\citep[e.g.][]{Mihosetal1995,Walkeretal1996} and major
\citep*[e.g.][]{Heyletal1996,Naabetal1999}. Yet, constructing realistic
equilibrium models is considerably difficult and in fact impossible
rigorously, i.e.\ requires some sort of approximation. 

\cite{Barnes1988} introduced a ``rather ad-hoc'' method for
constructing $N$-body models for a multi-component galaxy. He begins
by constructing separate spherical equilibrium $N$-body
\cite{King1966} models for the halo and bulge, which are then
superposed and allowed to relax into a new equilibrium over several
dynamical times.  Next, the gravitational potential of the disc is
slowly imposed, causing both a flattening and a radial contraction of
halo and bulge. Finally the disc component is populated with particles
drawn from a Maxwellian distribution with the local mean and
dispersion in agreement with the Jeans equations.

\cite{Hernquist1993} introduced another method, which makes it
possible to specify the density and velocity profiles of the various
components in a straightforward way. In this method, the positions of
all the particles in all components are determined first. This can be
carried out easily since the desired density profiles are known. The
Jeans equations are then used to find the velocity dispersions of the
various components, with those for the halo being found under the
assumption that the potentials of disc and bulge can be approximated
by their spherical average. Finally, individual velocities for all
components are drawn from a locally Maxwellian distribution with the
appropriate mean and dispersion velocities.

\cite*{BoilyKroupa2001} extended this approach to include a
non-spherical halo, but maintained the approximation to a Maxwellian
distribution for the particle velocities. This method is not rigorous
and \cite{Hernquist1993} himself suggested that ``In the future, it
will likely be necessary to refine the basic approach as computer
hardware and software permit simulations with particle numbers
significantly in excess of those discussed here.'' (he used
$N=49\,152$ in his empirical tests).

\cite*{KazantzidisMagorrianMoore2004} compared the properties of
$N$-body haloes based on the Maxwellian approximation with those
derived from an exact distribution function found through
\cite{Eddington1916} inversion. As their study clearly demonstrates,
the Maxwellian approximation results in non-equilibrium effects and is
inappropriate, for example, when modelling the tidal stripping of
substructure in CDM haloes. Unfortunately, Eddington inversion only
works for spherically symmetric equilibria and cannot be
straightforwardly applied to the problem of finding the
self-consistent distribution function of a multi-component system.

Therefore, \cite{KuijkenDubinski1995} followed an alternative method,
later expanded by \cite{WidrowDubinski2005}.  They start from an
analytical ansatz for the equilibrium distribution function of each
component in isolation, written in terms of energy and angular
momentum. In the case of the disc, the distribution function is also a
function of the ``vertical energy'' $E_z \equiv \frac{1}{2}v_z^2 +
\Phi(R,z) - \Phi(R,0)$, which for orbits close to the disc plane is
approximately conserved. These distribution functions are found for
predetermined isolated density distributions.  In the case of
\cite{KuijkenDubinski1995} these were a \cite{King1966} model bulge, a
lowered Evans model \citep{KuijkenDubinski1994} halo and an
exponential disc. The combined model is then found by using these
distribution functions in the combined gravitational potential of the
entire model. The major problem with this approach is summed up by
\cite{WidrowDubinski2005} when they state that the combined model
constructed in this way ``may bare [\emph{sic}] little resemblance to
the corresponding isolated components, a situation which is cumbersome
for model building''. This approach is also rather inflexible.

In this study, we therefore return to using Eddington inversion (in
fact a generalisation thereof) for the near-spherical components but
an analytic distribution function for the disc. Our method is
detailed is Section~\ref{sec:Method}, while Section~\ref{sec:ictests}
presents extended tests of the robustness and stability of the
constructed $N$-body equilibria. Finally, Section~\ref{sec:iccon} sums
up and concludes.

\section{The new method} 
\label{sec:Method}
Our method bears some relation to that of \cite{Barnes1988}, in that
the non-spherically symmetric component (the disc), is grown
adiabatically within the $N$-body halo. The bulge is assumed to be
spherically symmetric, though it would be a relatively straightforward
step to include a non-spherical bulge in a similar way to the disc.
The process has three stages:

\begin{enumerate}
\item Creating an equilibrium initial $N$-body representation of the
  spherically symmetric components of the galaxy (halo and bulge) in
  the presence of an external potential field corresponding to the
  monopole (spherical average) of the desired disc potential.
\item Evolving the $N$-body system by growing the non-monopole
  components of the disc potential adiabatically in a $N$-body
  simulation, in order to allow halo and bulge particles time to relax
  into the new potential.
\item Replacing the external potential field representing the disc
  component with an $N$-body representation.
\end{enumerate}

\subsection{Creating the initial \boldmath $N$-body halo and bulge}
\label{sec:Method:sph}

We create an $N$-body realization of the spherically symmetric
components of the system using Cuddeford's (\citeyear{Cuddeford1991})
method, an extension of the model by \citep{Osipkov1979} and
\citep{Merritt1985}, which in turn extended Eddington's
(\citeyear{Eddington1916}) original inversion technique to obtain the
distribution function for a spherical system with an isotropic
velocity distribution.

Cuddeford considered distribution functions of the form
\begin{equation}
  \label{eq:OsMerplus}
  f(\mathcal{E},L) = L^{2\alpha}\,f_0(Q).
\end{equation}
Here, $Q\equiv\mathcal{E}-L^2/2r_a^2$ as for Osipkov-Merritt models
($\mathcal{E}\equiv\Psi -\frac{1}{2}\bv^2$, $\Psi$ denoting the
negative of the gravitational potential) with anisotropy radius $r_a$.
The parameter $\alpha$ is constrained to be greater than $-1$. As
Cuddeford showed, $f(Q)$ is related to the density by an Abel integral
equation, which can (under the assumption that $f(Q < 0) = 0$) 
be inverted to yield
\begin{equation} \label{eq:Cudf}
  f_0(Q) =\frac{\sin(n-\frac{1}{2}-\alpha)\pi}
  {\pi\,\lambda(\alpha)\,\eta(\alpha)}\frac{\mathrm{d}}{\mathrm{d}Q}
  \int_0^Q \frac{\mathrm{d}^n \rho_{\mathrm{red}}}{\mathrm{d}\Psi^n}
           \frac{\mathrm{d}\Psi}{(Q-\Psi)^{\alpha + 3/2 - n}},
\end{equation}
where $n$ is defined as the largest integer equal to or less than $\alpha 
+ \frac{3}{2}$, and where the ``reduced density'' is given as
\begin{equation} \label{eq:Cudrho}
  \rho_{\mathrm{red}} =\frac{(1+
    r^{2}/r_{a}^{2})^{\alpha+1}}{r^{2\alpha}}\rho,
\end{equation}
while
\begin{equation}
  \eta(\alpha)=\left\{
    \begin{array}{lc}
      (\alpha+\frac{1}{2})(\alpha-\frac{1}{2})\dots(\alpha+\frac{3}{2}-n),
      & \alpha>-\frac{1}{2} \\
      1 & -1<\alpha\leq-\frac{1}{2},
    \end{array}\right.
\end{equation}
and
\begin{equation} \label{eq:Cudlambda}
  \lambda(\alpha) = 2^{\alpha+3/2}\,\pi^{3/2}\,
  \frac{\Gamma(\alpha+1)}{\Gamma(\alpha+3/2)}.
\end{equation}

This distribution function produces a spherically symmetric system
with a velocity distribution such that
\begin{equation} \label{eq:CudBeta}
  \beta(r)\equiv1-\frac{\sigma^2_{\theta}}{\sigma^2_r} = \frac{r^2 -
    \alpha r_{a}^{2}}{r^2 + r_{a}^{2}}.
\end{equation}
In the case where $\alpha=0$ this model reduces to an Osipkov-Merritt
model. In the case where $r_a\rightarrow\infty$, the anisotropy of the
halo is the same at all radii, $\beta=-\alpha$.

This approach has the advantage that the distribution function is
exact for a spherically symmetric system, and thus remains in
equilibrium, maintaining its original density profile. At no point is
the assumption of a Maxwellian velocity distribution made.  The only
restrictions to this method are that $\Psi=\Psi(r)$ is a monotonic
function of radius (only); that $\rho=\rho(r)$ and that the solution
to equation~(\ref{eq:Cudf}) must be physical, i.e.\ $f_{0}(Q)\geq0$.
This allows us to use a wide range of different halo (or bulge)
density profiles.

It is extremely useful that the derivation of equation~(\ref{eq:Cudf})
does not make the assumption that the potential of the system is that
due to the density profile through the Poisson equation.  This means
that it is straightforward to generalise this approach to find the
distribution function of a spherically symmetric component of a
larger, spherically symmetric, system. The term $\rho_{\mathrm{red}}$
in equation~(\ref{eq:Cudrho}) is replaced by the reduced density
$\rho_{\mathrm{red},i}$ the reduced density of the component $i$; the
term $\Psi$ in equation~(\ref{eq:Cudrho}) always refers to the
potential of the \emph{entire} system.

In the model, the presence of the disc component is taken into account
when calculating the distribution function of the halo and bulge. It
is impossible to do this exactly using this method, as the disc is not
spherically symmetric. The best approximation available in this case
is to take the spherical average of the disc potential. Then
the distribution functions of the bulge and halo are found in the
total potential of the halo, bulge and fictitious spherically averaged
disc. This prevents the radial contraction seen in the models
constructed by \cite{Barnes1988}.

Finally, we shall briefly describe how we choose initial conditions
from a model with distribution function of the
form~(\ref{eq:OsMerplus}). First, we draw an initial position from the
density in the usual way by inverting the cumulative mass profile for
the radius. Secondly, we sample velocities by introducing
pseudo-elliptical coordinates $(u,\eta)$ in velocity space such that
$v_r=u\cos\eta$ and $v_t=u(1+r^2/r_a^2)^{-1/2} \sin\eta$. Then we pick
values for $\eta$ and $u$ at random from the distributions
$p(\eta)=\sin^{1+2a}\eta$ and
$p(u)=u^{2+2\alpha}f_0(\Psi-\frac{1}{2}u^2)$, using the rejection
method for the latter.

\subsection{Evolving the halo and bulge to adapt to the disc}
\label{sec:Method:grow}

The second stage of creating the model is to evolve $N$-body model of
halo and bulge so that it is in equilibrium with the full disc
potential, rather than with its spherical average. To this end we use
the $N$-body code \textsf{gyrfalcON}, which is based on Dehnen's
(\citeyear{Dehnen2000, Dehnen2002}) force solver \textsf{falcON}---
though the axisymmetry of the system means that a code based upon an
expansion in spherical harmonics about the origin \citep[such as the
so-called self-consistent field codes, e.g.][]{HernquistOstriker1992}
would be well suited to this purpose.

The full potential of the disc, $\Phi_{\mathrm{disc}}(\bx)$ is grown
from its spherical average $\Phi_{\mathrm{disc},0}(r)$ according to
the formula
\begin{equation}
  \label{eq:growth}
  \Phi(\bx,t)=\;\Phi_{\mathrm{disc},0}(r) +
  A(t) \left[\Phi_{\mathrm{disc}}(\bx) -
    \Phi_{\mathrm{disc},0}(r)\right],
\end{equation}
where $A(t)$ is a growth factor that goes from $0$ at $t=0$ to $1$
smoothly in a time-scale, $t_{\mathrm{grow}}$, far longer than the
dynamical time in the region of the disc. The halo and bulge are then
allowed further time to relax completely under the influence of the
disc potential.

This process causes changes to the density distribution of the halo.
The spherically averaged density distribution is largely unchanged, but
the halo and bulge are somewhat flattened, and the iso-density
contours of both the halo and bulge become oblate. The degree of
flattening is dependent on the details of the various components, a
typical example is given in Section~\ref{sec:ictests}.

This stage is by some distance the slowest in the creating of the
initial conditions. However it is still takes a relatively small
fraction of the total CPU time of almost any scientifically
interesting simulation. It should also be noted that up until this
point the velocity distribution of the disc has not been a factor in
the calculations, so the halo and bulge created by this process can be
re-used in simulations with identical disc density profiles, but 
different disc kinematics.

\subsection{Populating the disc} 
\label{sec:Method:disc}
We start from the assumption that the motion of disc particles
decouples into its components in the plane of the disc, and
perpendicular to it. That is, we assume that the planar and vertical
components of the total energy,
$E_\parallel\equiv\frac{1}{2}(v_R^2+v_\phi^2)+\Phi(R,0)$ and
$E_\perp\equiv\frac{1}{2}v_z^2+\Phi(R,z)-\Phi(R,0)$, are both
separately conserved. This assumption is usually excellent for orbits
near the disc, for which always $|z| \ll R$. The potential $\Phi$ here
is, of course, the total potential of the system, i.e.\ that of the
disc model plus that of the evolved $N$-body system. The first is
computed as in \cite{DehnenBinney1998}, while the latter is
approximated using a potential expansion (so-called SCF) method
whereby ensuring axial symmetry by taking only $m=0$ and even $l$
terms.

We make an ansatz for the disc distribution function, following the
approach of \cite{Dehnen1999:DF}. As in that study, we first consider
the simplest distribution function, i.e.\ that of a dynamically
completely cold disc, in which all particles are on circular orbits.
For a disc with density $\rho=\Sigma(R)\,\delta(z)$ ($\Sigma$ being the 
surface density), this can be written as \citep[e.g.][]{Dehnen1999:DF}
\begin{equation}
  \label{eq:colddisc}
  f_{\mathrm{cold}}(E_\parallel,L_z,z,v_z)= 
  \frac{\Omega(R_{L_z})\,\Sigma(R_{L_z})}
  {\pi \kappa(R_{L_z})}\,
  \delta(E_\parallel-E_{\mathrm{c}}(R_{L_z}))\,\delta(z)\,\delta(v_z),
\end{equation}
where $\Omega(R)$ and $\kappa(R)$ denote the angular and epicycle
frequency at radius $R$ and $E_{\mathrm{c}}(R)$ the energy of the
circular orbit through $R$, while $R_{L_z}$ is the radius of the
circular orbit with angular momentum $L_z$. In order to obtain a warm
disc distribution function, one may replace the $\delta$-functions in
(\ref{eq:colddisc}) by exponentials, resulting in a locally Maxwellian
velocity distribution \citep[e.g.][]{Shu1969,KuijkenDubinski1995}.

However, as demonstrated by \cite{Dehnen1999:DF}, it is better to use
the following alternative form for the cold-disc model before
``warming'' it up.
\begin{equation}
  \label{eq:colddisc:alt}
  f_{\mathrm{cold}}(E_\parallel,L_z,z,v_z)= 
  \frac{\Omega(R_{E_\parallel})\,
    \Sigma(R_{E_{\parallel}})}
  {\pi \kappa(R_{E_\parallel})}\,
  \delta\Big(\Omega(R_{E_\parallel})
  [L_z-L_{\mathrm{c}}(R_{E_\parallel})]\Big)\,\delta(z)\,\delta(v_z),
\end{equation}
where $L_{\mathrm{c}}(R)$ is the angular momentum of the circular
orbit with radius $R$ and $R_{E_\parallel}$ the radius of the circular
orbit with energy $E_\parallel$. Again, the warm disc distribution
function is obtained by replacing the $\delta$-functions with
exponentials (noting that $z=0,\;v_z=0$ means that $E_\perp=0$), giving
\begin{eqnarray}
  f_{\mathrm{disc}}(E_\parallel,E_\perp,L_z) &=&
  \frac{\Omega(R_{E_\parallel})\,
    \tilde{\Sigma}(R_{E_{\parallel}})}
  {(2\pi)^{3/2} \kappa(R_{E_\parallel})}
  \;
  \frac{1}{z_d\,\sigma_z(R_{E_\parallel})}
  \exp\left( - \frac {E_\perp}
    {\sigma^2_z(R_{E_\parallel})}
  \right)
  \nonumber \\[1ex]
  \label{eq:warmdisc}
  &\times&
  \frac{1}{\tilde{\sigma}^2_R(R_{E_\parallel})}
  \exp\left( \frac
    {\Omega(R_{E_\parallel}) [L_z-L_{\mathrm{c}}(R_{E_\parallel})]}
    {\tilde{\sigma}^2_R(R_{E_\parallel})}
  \right).
\end{eqnarray}
Here, $\tilde{\Sigma}(R)$ and $\tilde{\sigma}_R(R)$
are sought such that the true surface-density and
radial velocity-dispersion profiles of the $N$-body
representation are those desired, to within an appropriate degree of
accuracy, while the velocity dispersion in the vertical direction
$\sigma^2_z(R)=\pi\,G\,z_d\,\Sigma(R)$ with
disc scale height parameter $z_d$.

%
%
\cite{Dehnen1999:DF} argued that equation~(\ref{eq:warmdisc}) gives a
more useful distribution function than that of \cite{Shu1969}; he
pointed that the warming of a disc can be described as an exponential
in the radial action $J_R$. The radial action is more closely
approximated by $\Omega(R_{E_\parallel})
[L_{\mathrm{c}}(R_{E_\parallel}) - |L_z|]/ \kappa(R_{E_\parallel})$
than by $[E_\parallel-E_{\mathrm{c}}(R_{L_z})]/\kappa(R_{L_z})$
\citep{Dehnen1999:Epi}. More practically, this distribution function
has the advantage that the value of $R_{E_\parallel}$ is generally a
far better approximation to the mean radius of an orbit than
$R_{L_z}$, which ensures that $\Sigma(R)$ and $\sigma_R(R)$ closely
resemble $\tilde{\Sigma}(R)$ and $\tilde{\sigma}_R(R)$, respectively.
This choice of distribution function also extends to negative $L_z$,
unlike that of Shu, allowing for a tail of counter-rotating stars.

When sampling from this distribution function, we attempt to minimise
noise in the particle distribution by sampling points more regularly
than random in phase space. We accomplish this by sampling orbits from
the density distribution $\tilde{\Sigma}(R)$, then placing $1\ll
N_{\mathrm{sam}}\ll\,N_\mathrm{disc}$ particles at points on each orbit.
Finally, we iteratively adapt $\tilde{\Sigma}(R)$ and $\tilde{\sigma}_R(R)$
so that the actual surface density and velocity dispersion profile match
the target profiles as closely as possible. The procedure for sampling
the planar part of the phase-space positions is very similar to that
proposed in \cite{Dehnen1999:DF} and its details are as follows.

\begin{enumerate}
  \itemsep1ex
\item \label{step:R} Draw a radius from a thin disc with surface
  density $\tilde{\Sigma}(R)$ (initially $\tilde{\Sigma}=\Sigma$) and
  set $E_\parallel$ to the energy of the circular orbit at $R$.
\item \label{step:Lz}
  Draw a number $\xi\in(0,1)$ and determine
  \begin{equation}
    L_z = L_{\mathrm{c}}(R) + \ln\xi\,\tilde{\sigma}_R^2(R)/\Omega(R).
  \end{equation}
  If $L_z\not{\in}[-L_{\mathrm{c}}(R),\,L_{\mathrm{c}}(R)]$ go back to
  step~\ref{step:R}.
\item \label{step:orbit} Integrate the orbit with these values of
  $E_\parallel$ and $L_z$ for one radial period $T_R$, find the radial
  frequency $\omega_R$, and tabulate $R(t)$ and $\dot{R}(t)$ for that
  orbit. Evaluate the correction factor
  $g_{\mathrm{corr}}\equiv\kappa(R)/\omega_R$ 
  \citep[see][]{Dehnen1999:DF}.
\item \label{step:sample} Given $N_{\mathrm{orb}} \approx
  N_{\mathrm{disc}}^{1/2}$, choose $N_{\mathrm{sam}}$ to be either
  of the two integers next to $g_{\mathrm{corr}}N_{\mathrm{orb}}$ drawn
  with probabilities such that the mean equals $g_{\mathrm{corr}}
  N_{\mathrm{orb}}$. Next sample $N_{\mathrm{sam}}$ orbital phases
  $t_i\in(0,T_R)$ and azimuth angles $\phi_i\in(0,2\pi)$, and determine
  the corresponding phase-space points $\{R_i, \phi_i, \dot{R}_i,
  \dot{\phi}_i\}$ from the orbit.
\item \label{step:final} Repeat steps \ref{step:R} to
  \ref{step:sample} until a total of $N_{\mathrm{disc}}$ disc
  particles have been sampled.
\item \label{step:iterate} Finally, evaluate from the
  $N_{\mathrm{disc}}$ phase-space points the actual surface density and
  radial velocity dispersion, $\Sigma_{\mathrm{out}}$ and
  $\sigma_{\mathrm{out},R}$, of the $N$-body model and adapt
  \begin{equation}
    \tilde{\Sigma}(R) \to \tilde{\Sigma}(R)
    \, \frac{\Sigma(R)}{\Sigma_{\mathrm{out}}(R)},\qquad
    \tilde{\sigma}_R(R) \to \tilde{\sigma}_R(R)
    \, \frac{\sigma_R(R)}{\sigma_{\mathrm{out},R}(R)}.
  \end{equation}
  Then repeat the whole sampling procedure of steps \ref{step:R} to
  \ref{step:final} and iteratively adapt $\tilde{\Sigma}(R)$ and
  $\tilde{\sigma}_R(R)$ until no further improvement occurs.
\end{enumerate}

If we use quasi-random, rather than pseudo-random, numbers in this
procedure, it closely resembles the ``quiet start'' method
\citep[e.g.][]{Sellwood1987}. Once this is done we determine the
vertical component of each disc particle's position and velocity. We
assume that the local structure in the $z$-direction corresponds to that of
an isothermal sheet with a constant vertical scale height,
$z_\mathrm{d}$ \citep{Spitzer1942}. This leads to a density throughout
the disc $\rho(R,z)\propto {\textrm{sech}}^2(z/z_\mathrm{d})$.  $v_z$
is then drawn randomly from a normal distribution with $\sigma^2_z=\pi
G\Sigma(R_{E_\parallel})\,z_\mathrm{d}$.

\begin{figure}
  \centerline{\resizebox{\hsize}{!}{\includegraphics{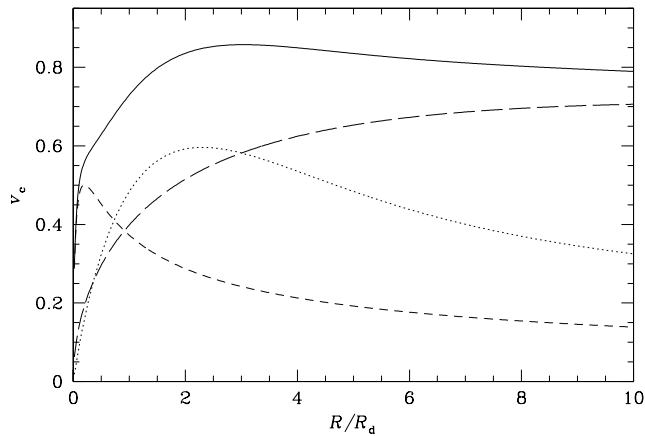}}}
  \caption{ Rotation curve for the test galaxy model.
    The solid line is the net rotation curve, also shown are the
    contributions from the disc (\emph{dotted}), bulge
    (\emph{short-dashed}), and halo (\emph{long-dashed}).
   \label{fig:rotcurve}
 }
\end{figure}

\section{Testing the new method}\label{sec:ictests}
We test our approach using a model based loosely upon the Milky Way as
modelled by \cite*{KlypinZhaoSomerville2002}. The disc is defined as
having an exponential surface density profile with a vertical
structure modelled by isothermal sheets, i.e.
\begin{equation} \label{eq:discrho} \rho_{\mathrm d}(R,z) =
  \frac{M_{\mathrm d}}{4\pi R_{\mathrm d}^2 z_{\mathrm d}}
  \,\exp\left(-\frac{R}{R_{\mathrm d}}\right) 
  \,\mathrm{sech}^2\left(\frac{z}{z_{\mathrm d}}\right),
\end{equation}
where $M_{\mathrm d}$ is the total disc mass, $R_{\mathrm{d}}$ is the
disc scale radius, and $z_{\mathrm d}$ is a scale height (though it
should be noted that this is not the e-folding height). 
The halo model is a truncated \citep[][hereafter NFW]{NFW1997} halo
with initially spherical density distribution
\begin{equation} \label{eq:tNFW} \rho_{\mathrm h} (r) =
  \rho_{\mathrm{c}}\,
  \frac{\textrm{sech}(r/r_{\mathrm{t}})}
  {(r/r_{\mathrm h})(1 + r/r_{\mathrm h})^{2}},
\end{equation}
while the bulge is modelled with a \cite{Hernquist1990} density profile
\begin{equation} \label{eq:rhohern}
  \rho_{\mathrm b} (r) = \frac{M_{\mathrm{b}}\,r_{\mathrm{b}}}{2\pi
    r(r_{\mathrm{b}} + r)^{3}}.
\end{equation}
We choose units such that Newton's constant of gravity $G=1$,
$R_{\mathrm d}=1$, and $M_{\mathrm d}=1$. We take the disc scale height
$z_{\mathrm d}=0.1$, the bulge mass $M_{\mathrm b}=0.2$ and scale length
$r_{\mathrm b}=0.2$. Scaling these values to the Milky Way, taking
$R_{\mathrm{d}}=3.5\,$kpc,
$M_{\mathrm{d}}+M_{\mathrm{b}}=5\times10^{10}\,\mathrm{M}_\odot$
\citep[as in][]{KlypinZhaoSomerville2002} gives a time unit
$\simeq14\,$Myr, and thus a velocity unit of $\sim250\,$km$\,$s$^{-1}$.

For the halo we take a scale radius $r_{\mathrm h}=6$, truncation radius
$r_{\mathrm t}=60$, and mass $M_{\mathrm h} = 24$, 79\% of which is
within the truncation radius.  The rotation curve for this model (with
velocities given in code units) is shown in Figure~\ref{fig:rotcurve}.

In all tests in which they were populated, the halo had $1\,200\,000$
particles, the bulge $40\,000$ and the disc $200\,000$. This corresponds
to each halo particle being 4 times more massive than a stellar
particle. We use gravitational softening lengths of $\epsilon=0.02$ for
disc and bulge particles, and $0.04$ for halo particles. This choice
ensures that the maximum force exerted by a single particle ($\propto
m_i/\epsilon_i^2$) is the same for all particles.

Simulations were performed with \textsf{gyrfalcON} with a minimum
time-step of $2^{-7}$, and a block-step scheme with largest time step
$2^{-4}$.  Individual particle time-steps were adjusted such that on
average the time-step
\begin{equation}
  \tau_i = \min\left\{\frac{0.01}{|\ba_i|},\;
    \frac{0.05}{|\Phi_i|}\right\},
\end{equation}
with $\Phi_i$ and $\ba_i$ the gravitational potential and acceleration
of the $i$th body in simulation units.

We first tested that the spherically symmetric components remain in
equilibrium in the spherically-symmetrised potential. This was tested
with various different choices for $r_a$ and $\alpha$
(equation~(\ref{eq:OsMerplus})) in the halo, in order to ensure that
Cuddeford inversion had been implemented correctly. The halo and bulge
profiles remained consistent to within a few softening lengths of the
centre, where softening and two-body relaxation have some small effect.
This testing was also used to assess whether the time integration
parameters were appropriate. Since energy was typically conserved to
within $0.05\%$ over 200 time units in these test simulations, we assume
that the time integration is sufficiently accurate. For all further
tests we only consider the isotropic case ($\alpha=0$,
$r_a\rightarrow\infty$, so $\beta=0$ throughout the halo).

\subsection{Testing the disc growth}\label{sec:dg}
The next step was to ensure that the growth of the full disc potential
from its monopole occurs without significant effect upon the radial
density profile of the halo or bulge, or upon their kinematics, and to
quantify the effect upon the shape of the resultant density
distribution. The disc potential is grown as per
equation~(\ref{eq:growth}), with a total growth time
$t_{\mathrm{grow}}=40$.  This is substantially longer than the dynamical
time $t_{\mathrm{dyn}}$ in the vicinity of the disc (for instance at
$R=3$, $t_{\mathrm{dyn}}\sim6$).

\begin{figure*}
  \centerline{
    \resizebox{86mm}{!}{\includegraphics{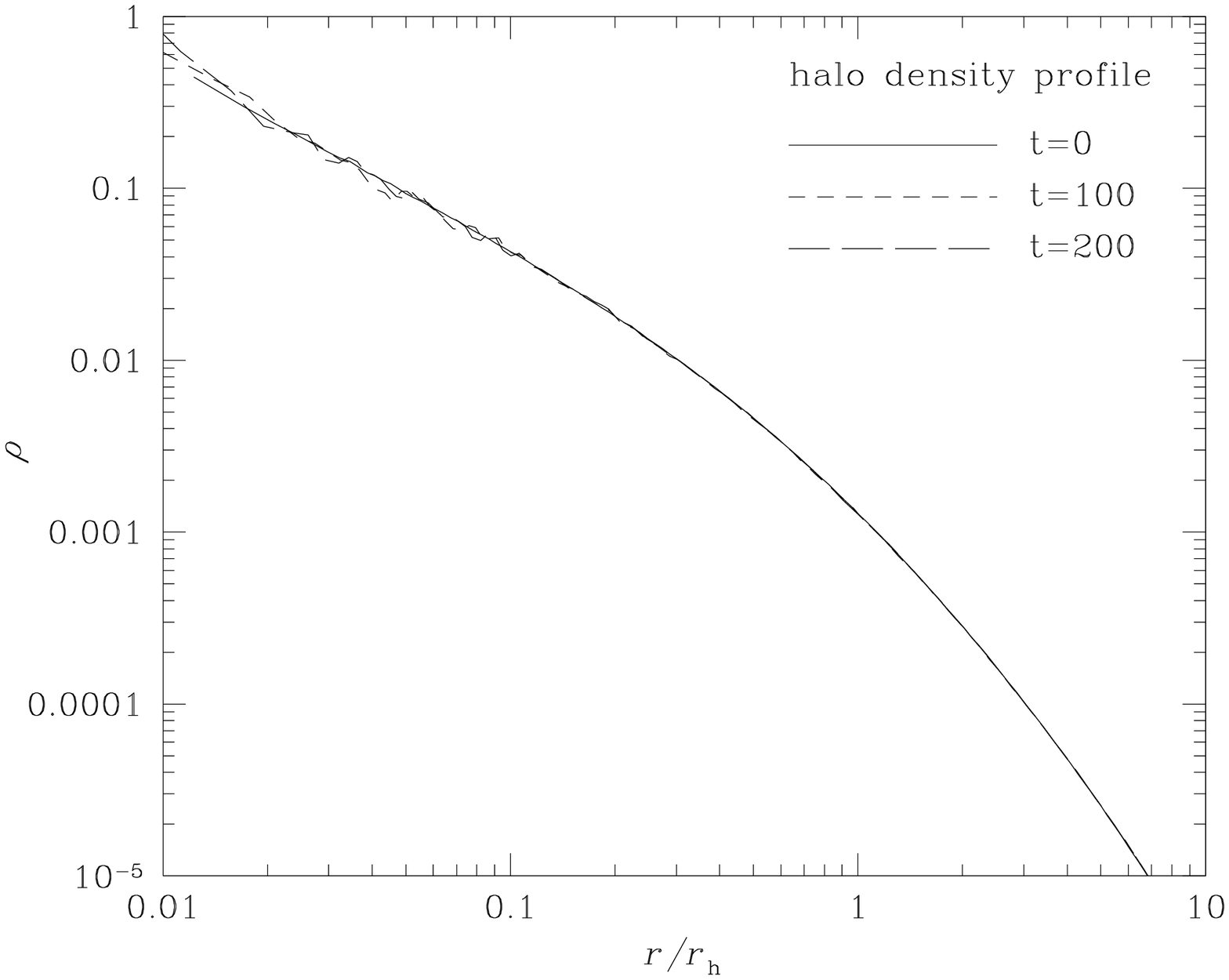}}
    \hfill
    \resizebox{86mm}{!}{\includegraphics{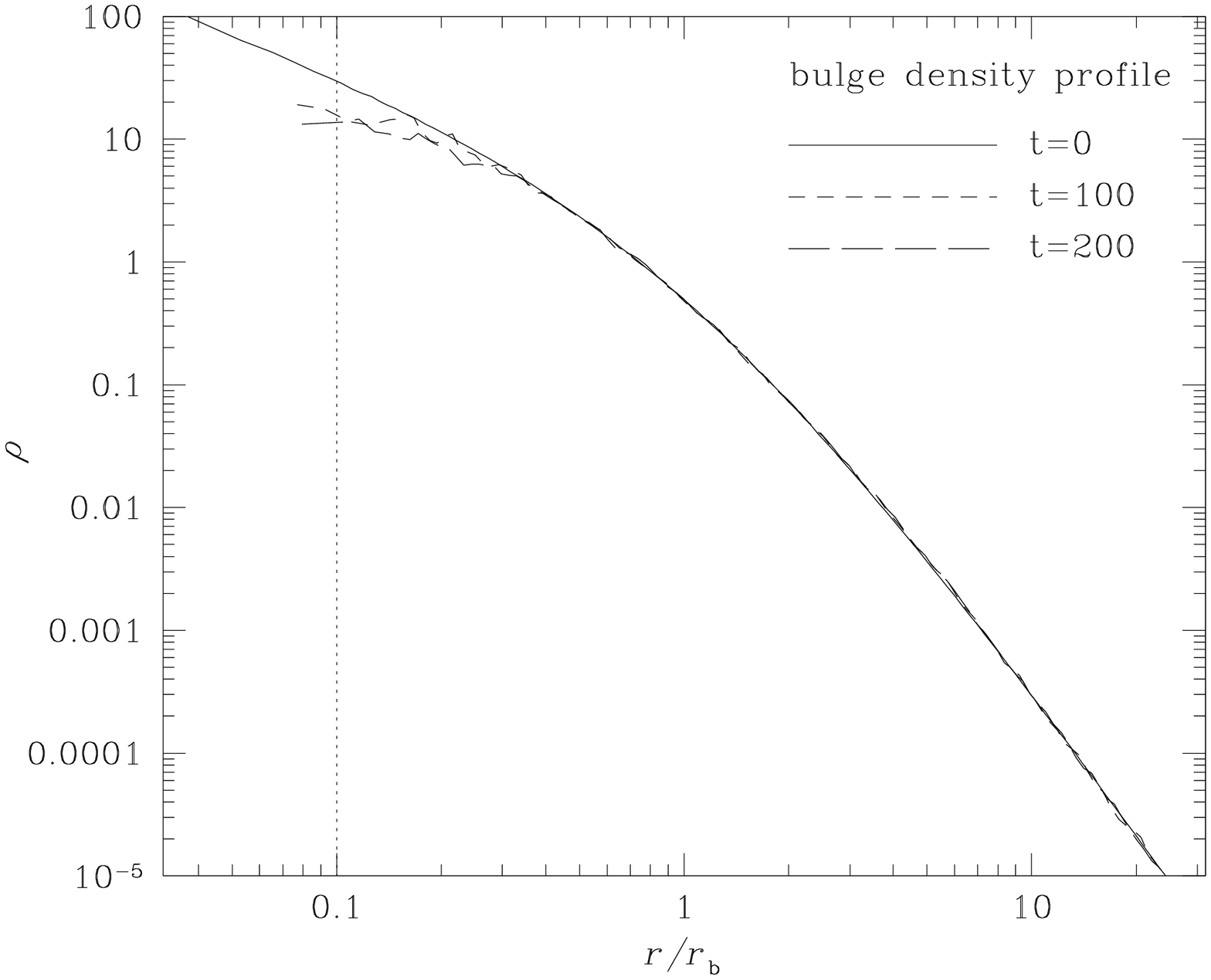}}
  }
  \caption{
    Spherically averaged density profile of the halo (\emph{left})
    and bulge (\emph{right}), shown just before growth of the full disc
    potential ($t=0$), and at 100 and 200 time units.
    The disc component was not populated in this simulation. The 
    density profiles remain more or less unchanged by the
    growth of the full potential, except at radii only slightly
    larger than the respective softening length
    ($\epsilon_{\mathrm{h}}=0.0066r_{\mathrm{h}}$ and 
    $\epsilon_{\mathrm{b}}=0.1r_{\mathrm{b}}$, indicated by
    the dotted vertical line).
    \label{fig:dg:hbpro}
  }
\end{figure*}
Figure~\ref{fig:dg:hbpro} shows the evolution of the density profiles of
halo and bulge. Both are maintained to within a high degree of accuracy
through the growth of the full potential, and then remain close to
unchanged for a long period of time. The halo density is slightly raised
in the inner $\sim0.02r_{\mathrm h}=0.12$ and the bulge density is
slightly lowered in the inner $\sim0.3r_{\mathrm b}=0.06$, in each case
corresponding to only a few softening lengths. This is approximately the
same as observed in simulations with no disc growth and is likely caused
by mass segregation due to (artificial) two-body relaxation.

\begin{figure*}
  \centerline{
    \resizebox{86mm}{!}{\includegraphics{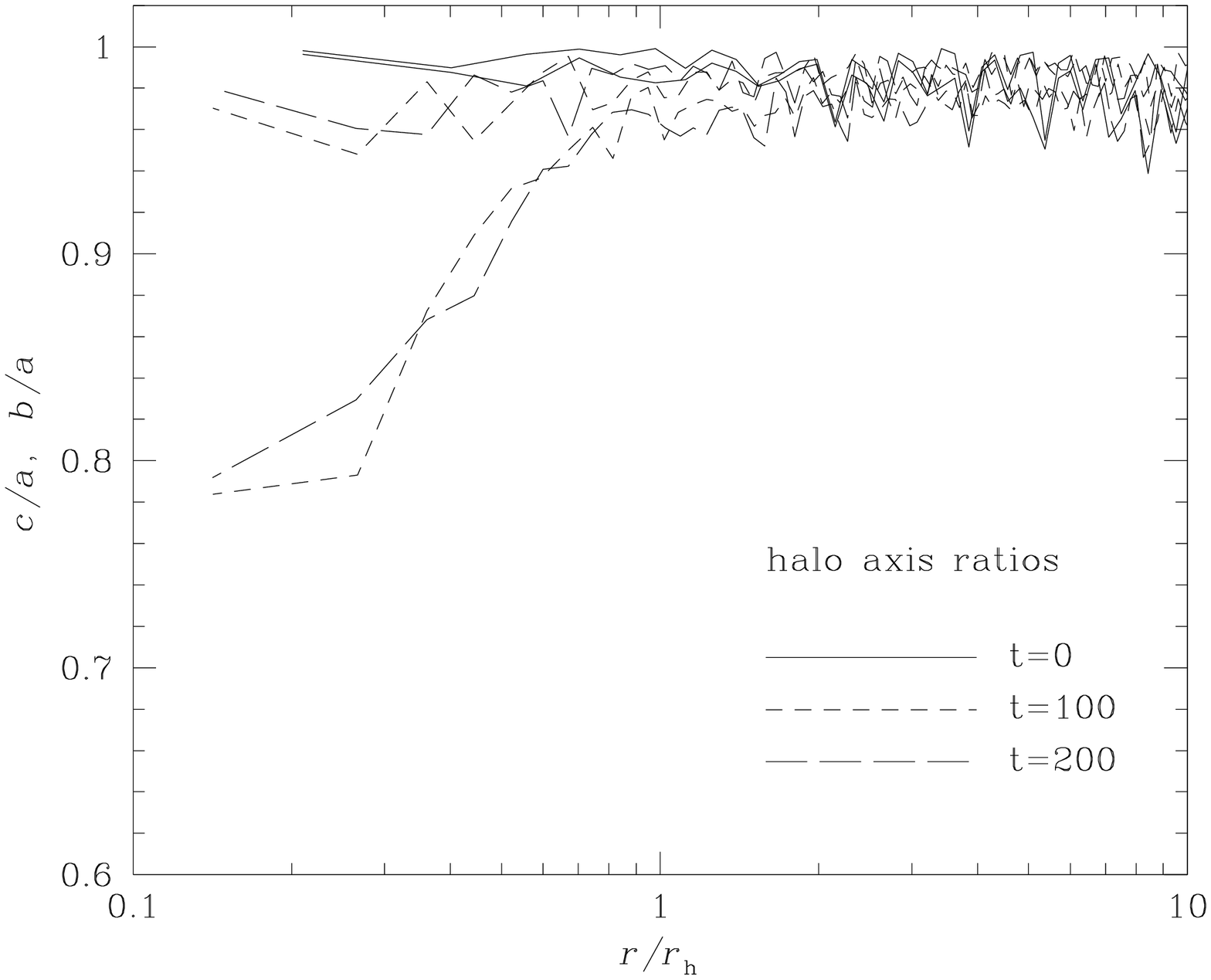}}
    \hfill
    \resizebox{86mm}{!}{\includegraphics{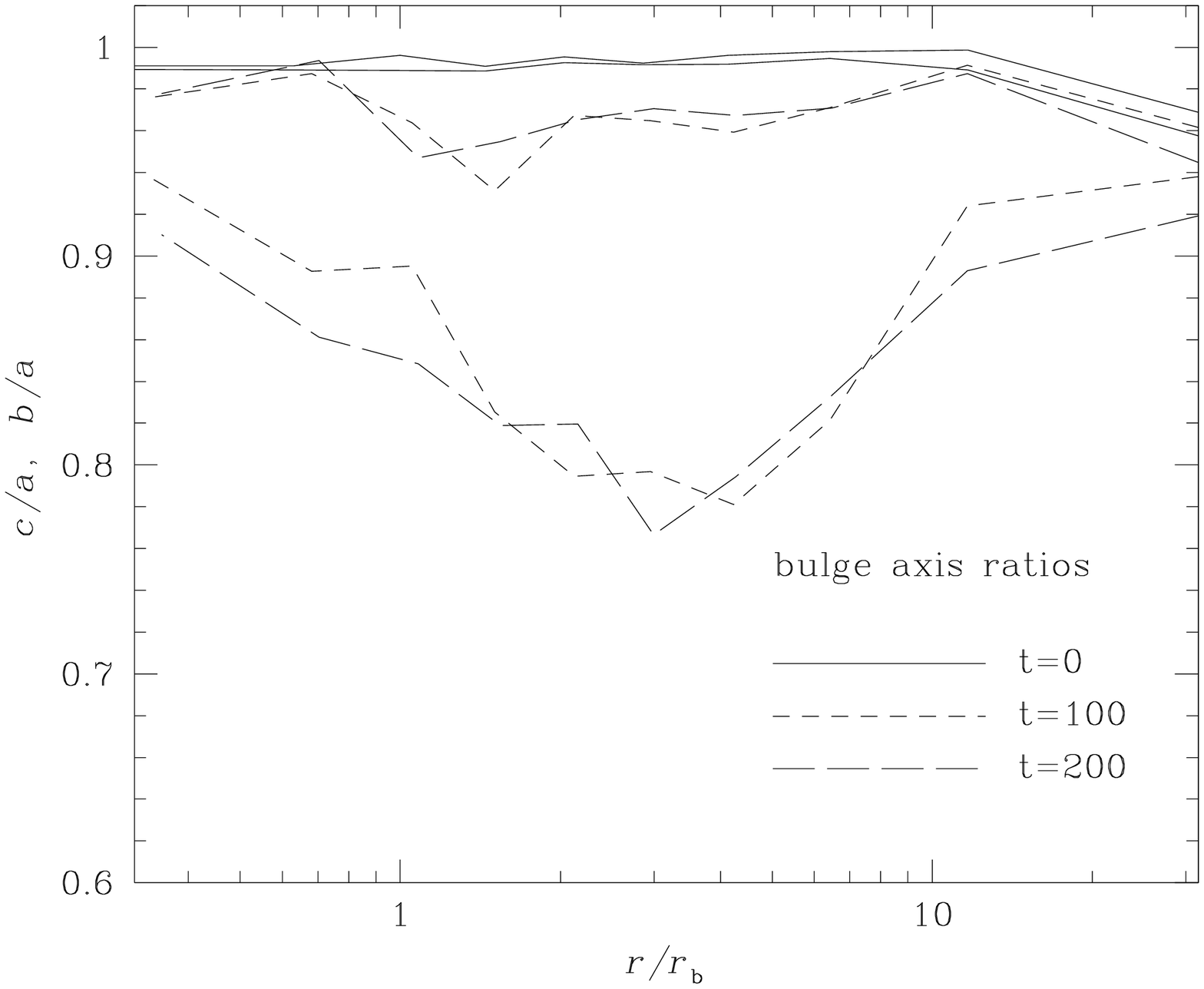}}
  }
  \caption{Minor and intermediate axis ratios ($c/a\,<\,b/a$
    respectively), for both the halo (\emph{left}) and bulge
    (\emph{right}) in the same simulations as in
    Fig.~\ref{fig:dg:hbpro}. Particles are binned in density, and the
    values plotted here are the axis ratios of these bins against their
    median radius. Deviations from sphericity ($c/a=b/a=1$)
    at $t=0$ are entirely due to discreteness noise.
    \label{fig:dg:shape}
  }       
\end{figure*}

To quantify the flattening of halo and bulge due to the growth of the
full disc potential, we determine the axis ratios of the halo and bulge
as a function of radius. To this end, we first estimate the local
density at each particle's position by the nearest-neighbour method of
\cite{CasertanoHut1985}, using the 15 nearest neighbours (of the same
component). Next, we bin the particles in estimated density and evaluate
the axis ratios in each bin as the square roots of the ratios of the
eigenvalues of the moment-of-inertia tensor\footnote{In the literature
  one often finds axis ratios computed for bins in spherical radius
  \citep[e.g.][]{Kazanzidisetal2004}. This not only requires accurate
  knowledge of the centre position, but more importantly results in a
  substantial bias (because of the usage of spherical symmetry). For a
  triaxial body, for example, the axis ratios at small radii are
  drastically overestimated \citep{Athanassoula2007}.
  Binning in potential energy gives somewhat better but still
  unsatisfactory results (since the gravitational potential is less
  flattened than the density).}. Figure~\ref{fig:dg:shape} plots the
axis ratios such obtained for halo and bulge versus the radius (the
median radius of the corresponding density bin).
Obviously, there is some small effect on the bulge and halo shape from the
growth of the disc. In both cases the shortest ($c$) axis is very close
to the $z$-axis as expected. The degree of flattening is relatively
modest, with $c/a\simeq0.8$ in the most flattened shells. In the case of
the halo this is within the inner $\sim 0.3 r_{\mathrm h} =
1.8R_{\mathrm d}$. In the case of the bulge the most flattened shells
are at $\sim 3 r_{\mathrm b} = 0.6R_{\mathrm d}$. The bulge is less
flattened at smaller radii because the disc is of finite thickness
(scale height $z_{\mathrm d}=0.1R_{\mathrm d}$), which means that the
full potential is less flattened -- when compared to the spherical
average -- in the inner parts of the bulge than in the outer. The same
effect would be observable in the halo with sufficient resolution.

\begin{figure}
  \centerline{\resizebox{\columnwidth}{!}{\includegraphics{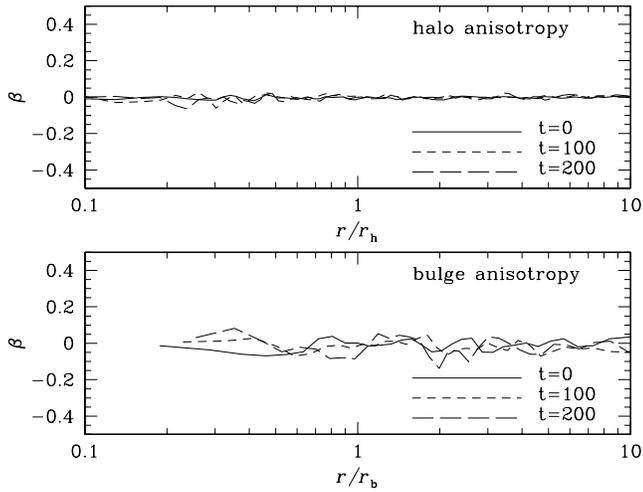}}}
  \caption{Anisotropy parameter $\beta$, determined from
    spherical shells, plotted against radius for the halo and bulge
    in the same simulations as in Fig.~\ref{fig:dg:hbpro}.
    Initial conditions were defined as having $\beta=0$ and 
    variation from that at $t=0$ is due to discreteness
    noise. 
    \label{fig:dg:beta}
  }
\end{figure}
For both halo and bulge Figure~\ref{fig:dg:beta} shows the evolution of
the radial profile of the velocity anisotropy parameter $\beta$ as
determined from spherical shells. Both components are initially
isotropic and, to within numerical accuracy, remain so during and well
after the growth of the full disc potential.

\subsection{Testing the disc model}\label{sec:solodisc}

Before testing the full compound galaxy, we test the disc model in
isolation, i.e.\ we follow the evolution of the populated disc
component in the static external potential $\Phi(R,z)$, the
cylindrically-symmetrised potential in which the distribution function
of the disc was constructed. The purpose of this test is to ensure that
the approximate nature of our disc distribution function (the assumption
that the planar and vertical energies are separately conserved) does not
compromise our approach, i.e.\ that the disc $N$-body model is close to
equilibrium. 

The main difficulty here is that we want the disc model to be stable to
axisymmetric perturbations, but not to non-axisymmetric instabilities,
such as spiral waves and bar modes, both of which one would expect to
see. These instabilities cause redistribution of the mass of the disc in
both the $R$ and $z$ directions \citep[e.g.][]{Hohl1971,AM2002}. In the
tests presented here, we prevent non-axisymmetric modes from growing by
a technique pioneered by Athanassoula (private communication): the
azimuth of every disc particle is randomised after every block-step,
thus destroying any coherent non-axisymmetric perturbation.

The distribution function defined by equation~(\ref{eq:warmdisc}) allows
for many possible $\sigma_R(R)$. In practise our current implementation
restricts the possibilities to either $\sigma_R\propto
\exp(-R/R_\sigma)$, or $\sigma_R$ is such that the \cite{Toomre1964}
stability parameter
\begin{equation}
  \label{eq:Q}
  Q \equiv \frac{\sigma_R\,\kappa}{3.36\,G\Sigma}
\end{equation}
is constant throughout the disc (this should not be confused with the
Osipkov-Merritt $Q$ in equation~(\ref{eq:OsMerplus})). A stellar disc
is known to be unstable to axisymmetric waves if Toomre's $Q<1$.

\begin{figure}
  \centerline{\resizebox{\hsize}{!}{\includegraphics{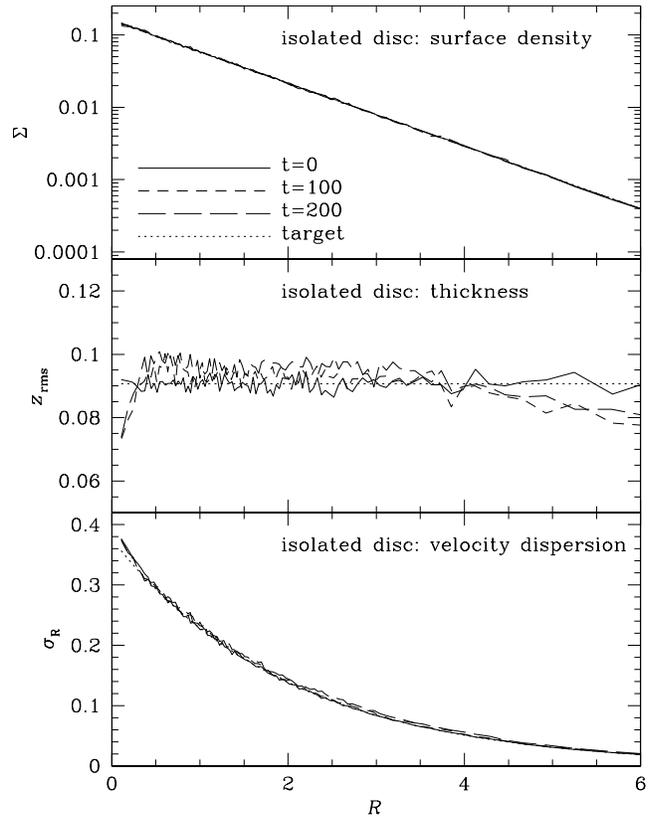}}}
  \caption{Radial profiles of surface density (\emph{top}), thickness
    (\emph{middle}), and radial velocity dispersion (\emph{bottom})
    for an isolated disc model. The randomising-azimuth method has
    been used to prevent the growth of bar or spiral modes.
    \label{fig:ld}
  }
\end{figure}

We test two models, one with constant $Q=1.2$, and one with
$\sigma_R\propto \exp(-R/2R_{\mathrm{d}})\;$ (i.e.\
$R_\sigma=R_{\mathrm{d}}/2$ so that $\sigma^2_R\propto\Sigma$), with the
constant of proportionality defined such that $Q(R=R_\sigma)=1.2$. The
two give qualitatively similar results, so only the results from the
latter distribution function are shown in Fig.~\ref{fig:ld} for
simplicity.
The surface density of the disc model is preserved to within the inner
$\sim0.1-0.2R_{\mathrm{d}}$, and out to beyond $6R_{\mathrm{d}}$. The
radial velocity dispersion $\sigma_R$ is preserved over a similar range.

In the middle panel of Fig.~\ref{fig:ld}, we plot the r.m.s.\ value of
$z$ as a measure of the disc thickness. For a disc with
$\rho\propto\mathrm{sech}^2(z/z_{\mathrm{d}})$ one expects
$z_{\mathrm{rms}} = \pi z_\mathrm{d}/\sqrt{12}
\approx0.907z_\mathrm{d}$, indicated by a dotted line. The disc
thickness remains near constant in the range $0.3\lesssim R \lesssim 4$,
with a slight warming which can reasonably be attributed to particle
softening.  The disc becomes somewhat flattened in its very inner 
($\sim0.3R_{\mathrm{d}}$) and outer ($R>4R_{\mathrm{d}}$) parts. 
This is presumably caused by too low initial $z$-velocities, which were
assigned assuming that the vertical force is dominated by the local
disc. In the inner parts of the disc, the bulge has a non-negligible
contribution to the vertical force field, and in the outer parts 
the local disc is very tenuous, so the contribution of the 
halo (and the monopole of the disc) to the vertical force field is
significant. It should also be noted that the approximation
that the planar and vertical motion decouple is worst in the inner
parts of the disk, which may well contribute to the flattening there. 

\subsection{Testing the full model}

\subsubsection{A constrained model}
Finally we test the complete compound model. First we wish to examine
the behaviour of the system in the absence of bar or spiral
instabilities. In order to do so, we utilise the same method of
randomising the azimuth (of disc particles only) as in
Section~\ref{sec:solodisc}. 

\begin{figure*}
  \centerline{
    \resizebox{86mm}{!}{\includegraphics{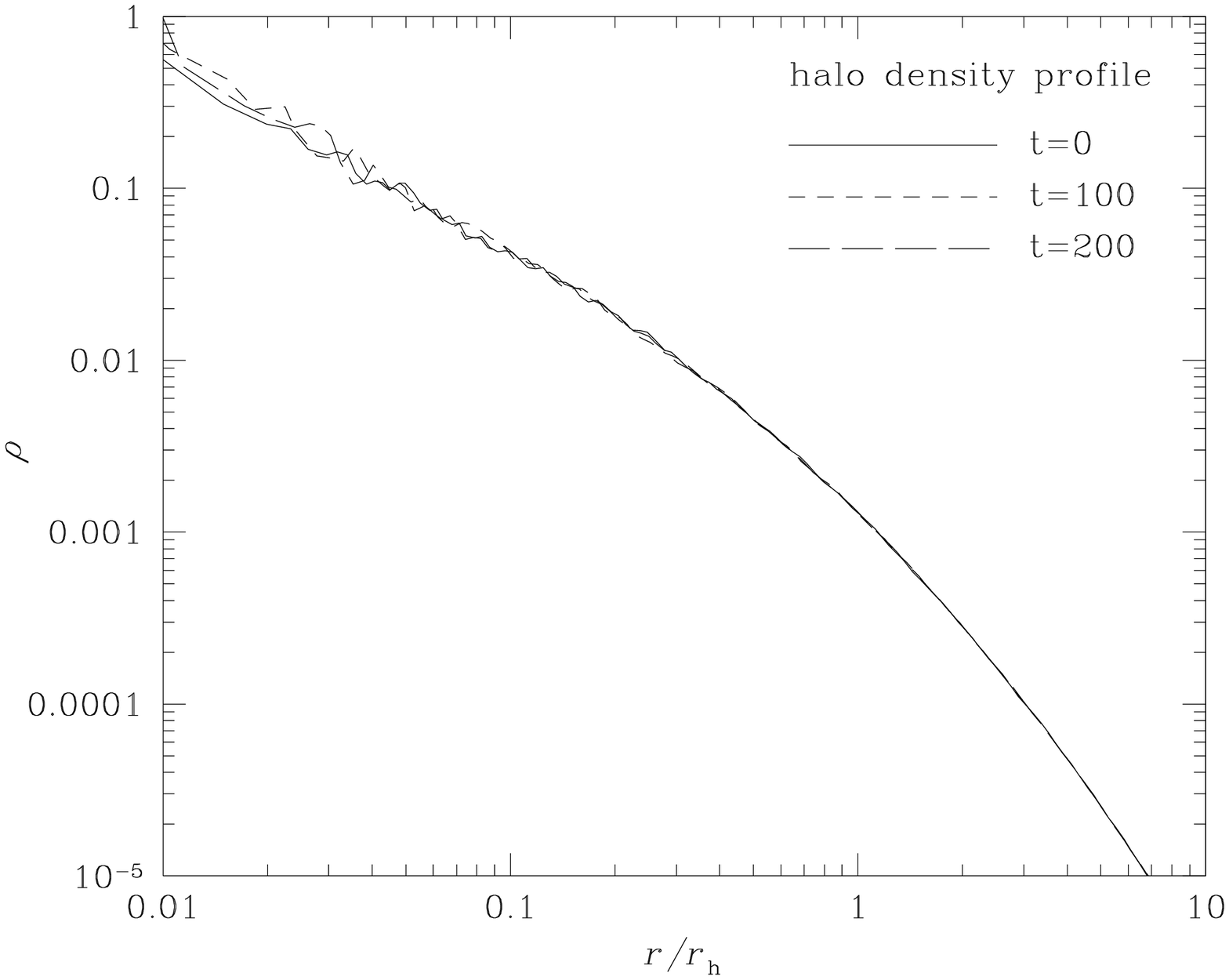}}
    \hfill
    \resizebox{86mm}{!}{\includegraphics{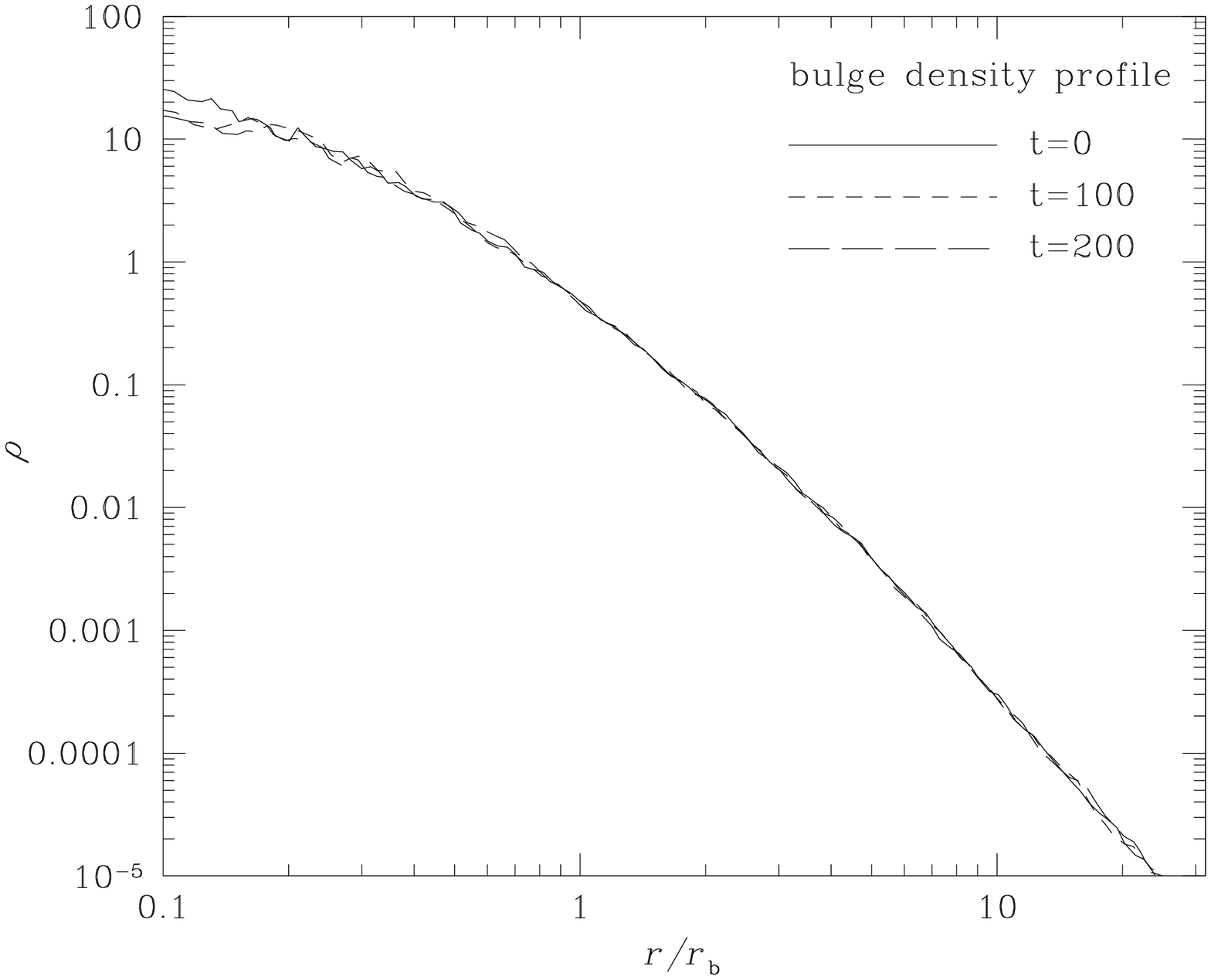}}
  }
  \caption{
    Spherically averaged density profile of the halo (\emph{left})
    and bulge (\emph{right}) in a fully populated simulation with
    non-axisymmetric perturbations suppressed (see text). 
    The profiles are shown at $t=0$, when the full $N$-body model
    was populated but before it has evolved; and at two times during the
    full $N$-body simulation. Symmetry about the 
    origin was enforced, to avoid numerical difficulties relating 
    to off-centring. 
    \label{fig:full:hbpro}
  }
\end{figure*}

However, using this method with a live halo and bulge is not
straightforward, since a live $N$-body simulation will often drift
slightly from its original centre. When this happens the disc will be
pulled in same the direction as the inner halo, so that randomising
azimuths around the origin would alter the model structure. In order to
prevent such a drift, we employed another symmetrisation method: the
distribution of particles in halo and bulge was kept reflection
symmetric about the origin at all times. To this end, we treated
particles in bulge and halo as pairs and enforced, initially and after
each time step (during both the disc growth and the fully populated
simulation), that for each pair positions and velocities both add up to
zero.

\begin{figure*}
  \centerline{
    \resizebox{86mm}{!}{\includegraphics{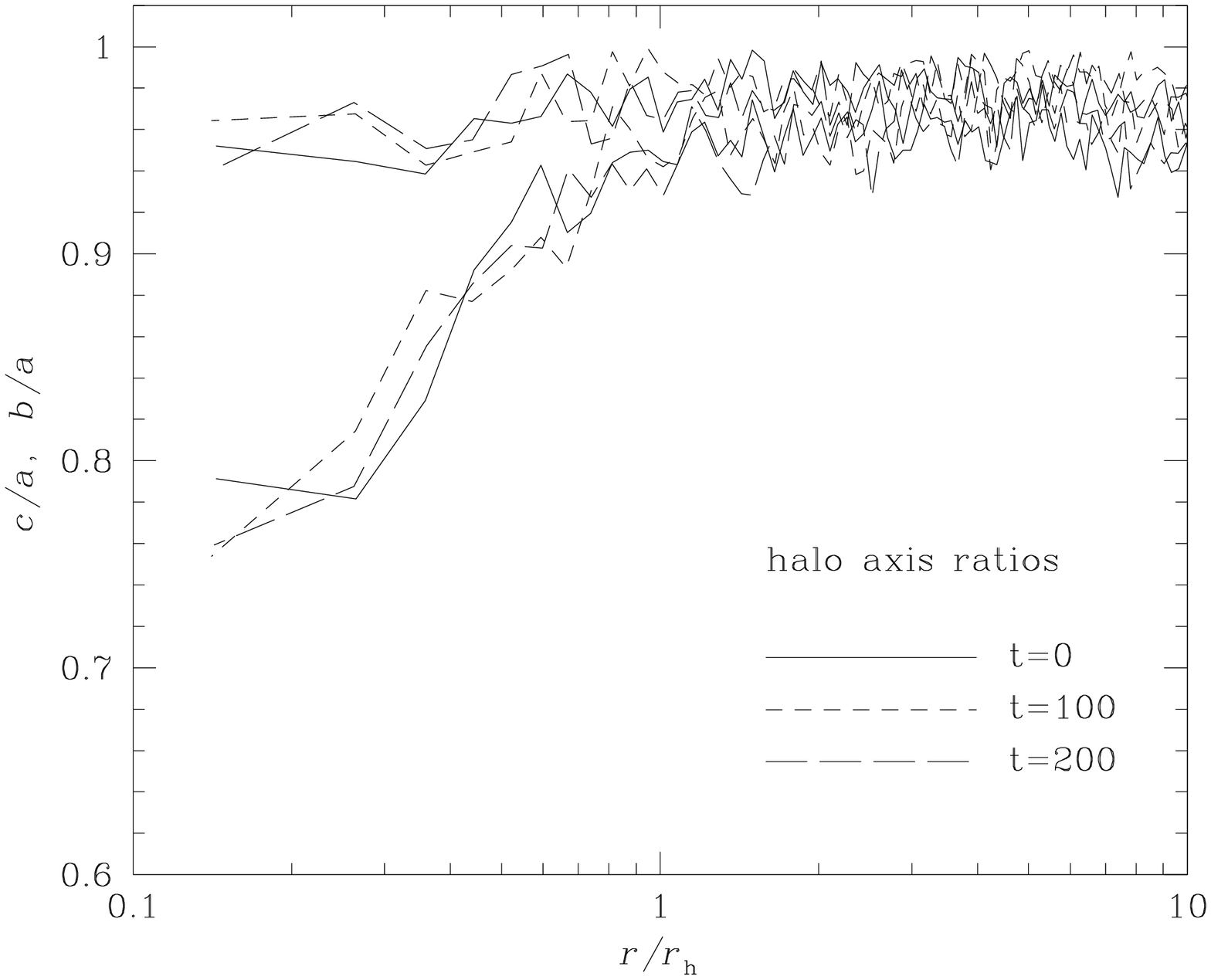}}
    \hfill
    \resizebox{86mm}{!}{\includegraphics{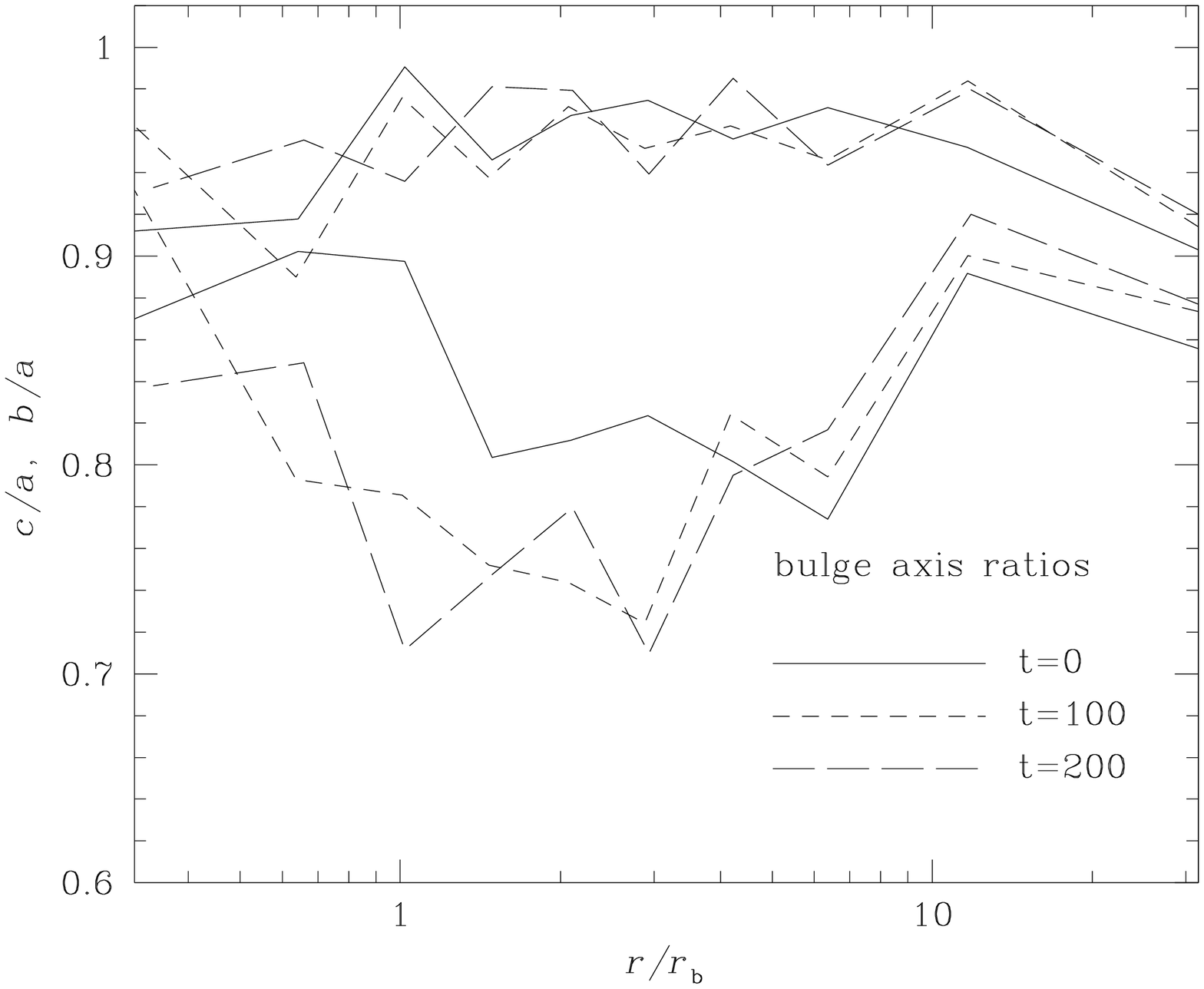}}
  }
  \caption{Minor and intermediate axis ratios ($c/a\,<\,b/a$
    respectively), for both the halo (\emph{left}) and bulge
    (\emph{right}) in the same simulations as in
    Fig.~\ref{fig:full:hbpro}. Particles are binned in density, and the
    values plotted here are the axis ratios of these bins against their
    median radius.
    \label{fig:full:shape}
  }       
\end{figure*}

\begin{figure}
  \centerline{\resizebox{\hsize}{!}{\includegraphics{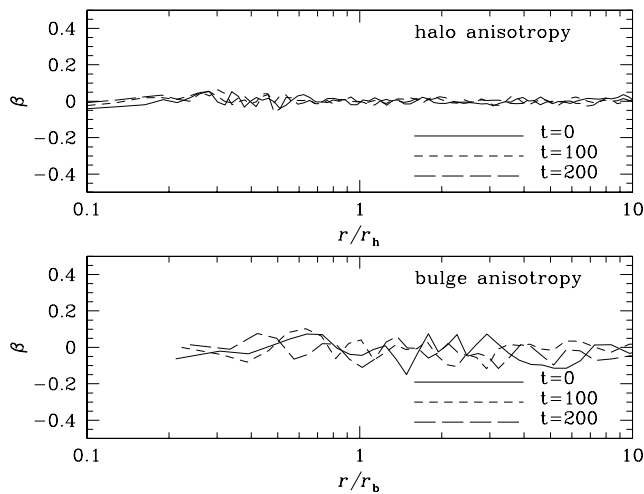}}}
  \caption{Anisotropy parameter $\beta$, as determined from spherical
    shells, vs.\ radius for the halo and bulge in the same 
    simulations as 
    Figs.~\ref{fig:full:hbpro} \& \ref{fig:full:shape}.
    \label{fig:full:beta}
  }
\end{figure}

\begin{figure}
  \centerline{\resizebox{\hsize}{!}{\includegraphics{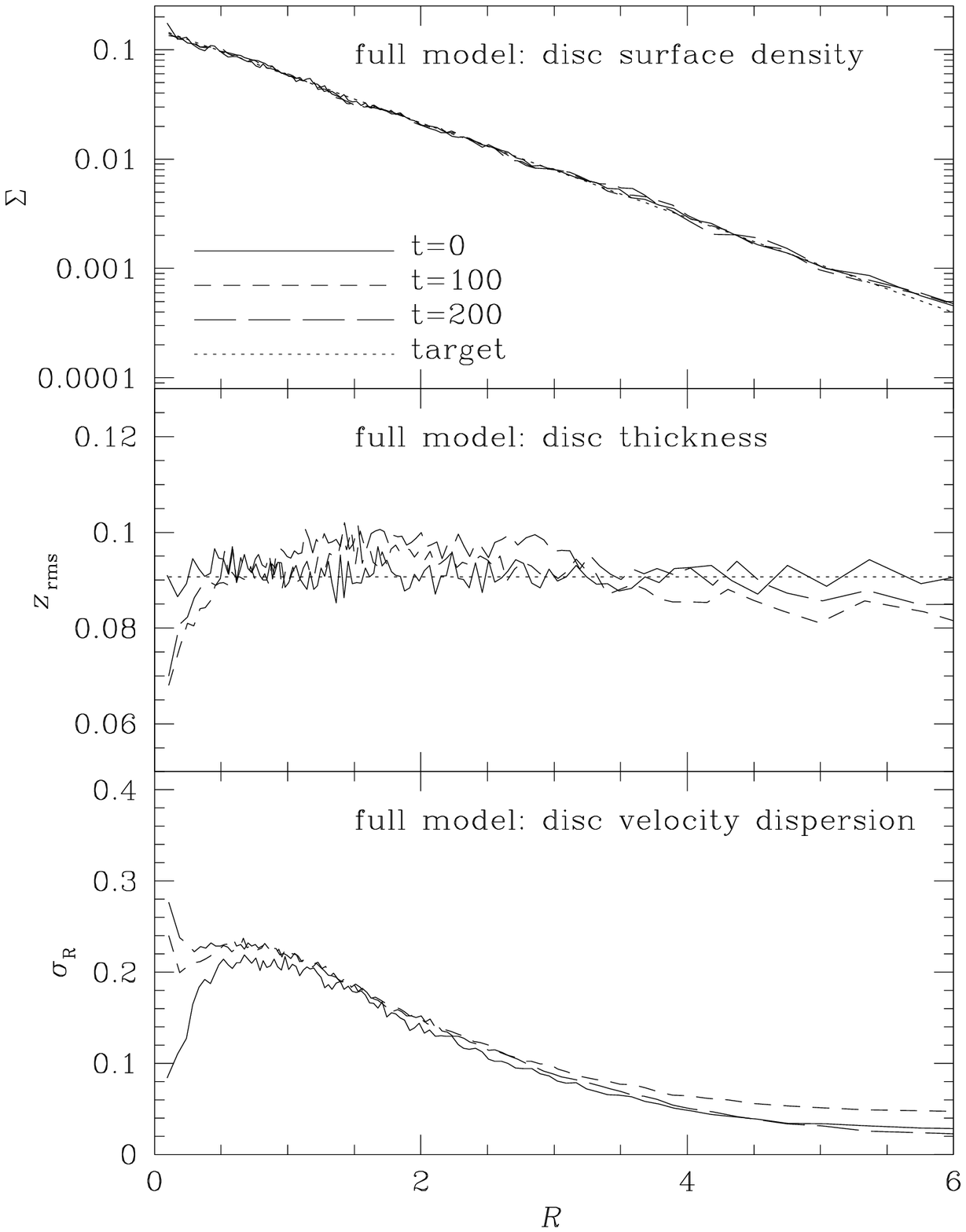}}}
  \caption{Radial profiles of surface density (\emph{top}), thickness
    (\emph{middle}), and radial velocity dispersion (\emph{bottom})
    of the disc component in a full $N$-body model, whose halo
    and bulge properties are shown in Figs.~\ref{fig:full:hbpro} to
    \ref{fig:full:beta}. The randomising-azimuth method has been used
    to prevent the growth of bar or spiral modes (see text).
    \label{fig:full:ld}
  }
\end{figure}

In this test we set the velocity dispersion $\sigma_R$ of the disc such
that Toomre's $Q=1.2$ at all radii. The full disc potential was grown
over a period $t_{\mathrm{grow}}=40$, as in Section~\ref{sec:dg}, and
then kept in place for a time $t_{\mathrm{hold}}=20$ to ensure that the
halo and bulge had fully relaxed in its presence. Only then was the disc
populated with particles. Simulations were then run to observe the
evolution over another 200 time-units. Times quoted in
Figs.~\ref{fig:full:hbpro}-\ref{fig:full:ld} take $t=0$ to be the time
when the fully populated simulation begins (\emph{not} the start of the
growth of the full disc potential, as in Section~\ref{sec:dg}).

Figures~\ref{fig:full:hbpro} and \ref{fig:full:shape} show that the
properties of the halo and bulge are maintained to within a high degree
of accuracy throughout the simulation. The halo density is slightly
raised in the inner $\sim0.02r_{\mathrm h}=0.12$ and the bulge density
is slightly lowered in the inner $\sim0.3r_{\mathrm b}=0.06$, much as
they were in simulations with an unpopulated disc
(Fig.~\ref{fig:dg:hbpro}). Both components remain isotropic (with some
noise, Fig.~\ref{fig:full:beta}), and while they are non-spherical, they
do not become significantly more or less aspherical over the course of
the simulation (Fig.~\ref{fig:full:shape}).

Figure~\ref{fig:full:ld} (top panel) shows that the surface density of
the disc is nearly unchanged at all radii throughout the entire
simulation. Also the disc thickness remains near constant over the full
radial range. The small deviations are similar to those observed for the
disc in isolation (Fig.~\ref{fig:ld}). In particular, the disc does not
thicken up as one might expect from heating due to collisions with (more
massive) interloping halo particles. This is probably a side effect of
the randomisation of the azimuthal angle of the disc particles. This
randomisation has the effect that disc particles are rarely close to any
halo particle crossing the disc for more than one time step.

The radial velocity dispersion $\sigma_R$ in the disc
(Fig.~\ref{fig:full:ld}, bottom panel) is nearly unchanged in the range
$0.5<R<3$. In the outer parts of the disc there is some variation, but
no consistent warming or cooling. In the inner parts of the disc there
is a clear increase in $\sigma_R$. This warming clearly does not affect
the density distribution. It can reasonably be attributed to the facts
that the decoupling of the vertical and planar motion is a poor
approximation in the inner disc, and that, while the choice of a
distribution function with constant $Q$ can be useful in that it avoids
the inner parts of the disc being very hot, it does lead to the velocity
dispersion being unrealistically low at the very inner radii.

\subsubsection{Unconstrained tests}

Finally we run simulations in which the galaxy model is unconstrained 
from the moment the disc component is populated. This means that 
the disc component is free to develop bar modes and other instabilities.

\begin{figure}
  \centerline{\resizebox{70mm}{!}{\includegraphics{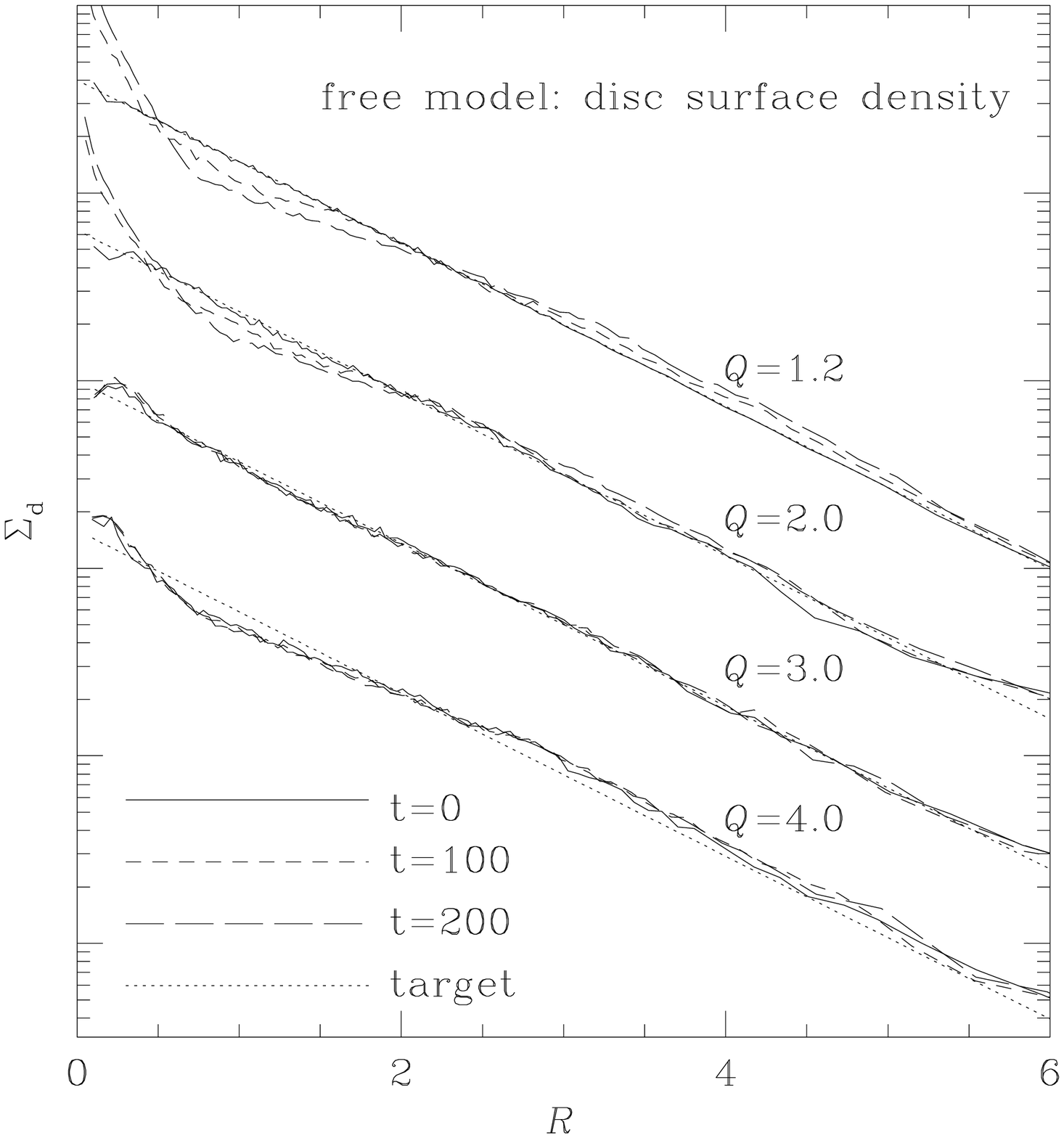}}}
  \caption{Surface density profiles for the discs in unconstrained
    simulations with four different values of Toomre's $Q$ initially.
    Results for the different simulations are offset by a constant.
    \label{fig:free:disc}
  }
\end{figure}
\begin{figure}
  \centerline{\resizebox{\hsize}{!}{\includegraphics{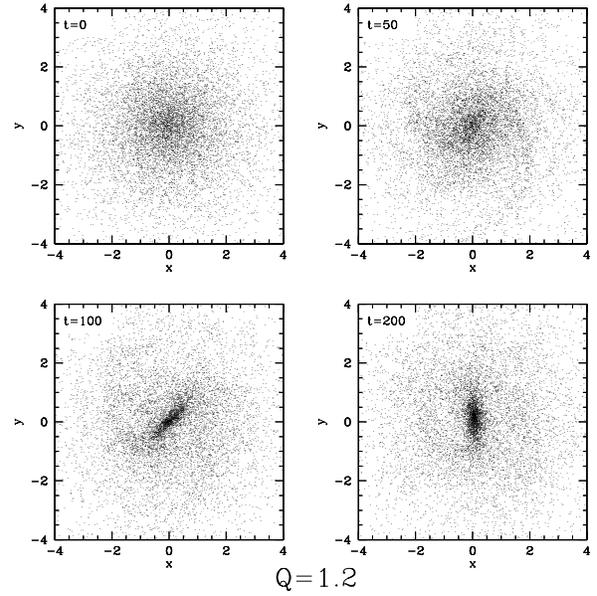}}}
  \caption{Scatter plot of disc particle positions in the $x$-$y$ plane
    at various times in an unconstrained simulation with initial
    $Q=1.2$.  Only 1 in every 20 particles is plotted, in the interests
    of clarity. \label{fig:full:scat12}
  }
\end{figure}
\begin{figure}
  \centerline{\resizebox{\hsize}{!}{\includegraphics{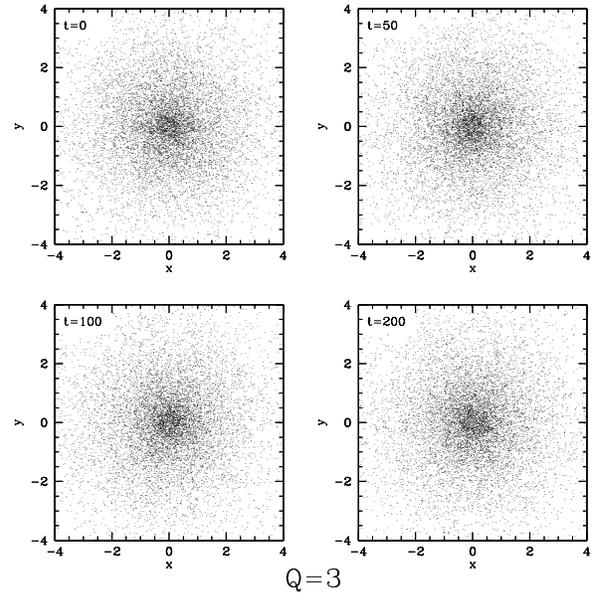}}}
  \caption{Scatter plot of disc particle positions in the $x$-$y$ plane
    at various times in an unconstrained simulation with initial $Q=3$.
    As in figure~\ref{fig:full:scat12}, only 1 in every 20 particles 
    is plotted. \label{fig:full:scat3}
  }
\end{figure}
\begin{figure*}
  \centerline{\resizebox{130mm}{!}{\includegraphics{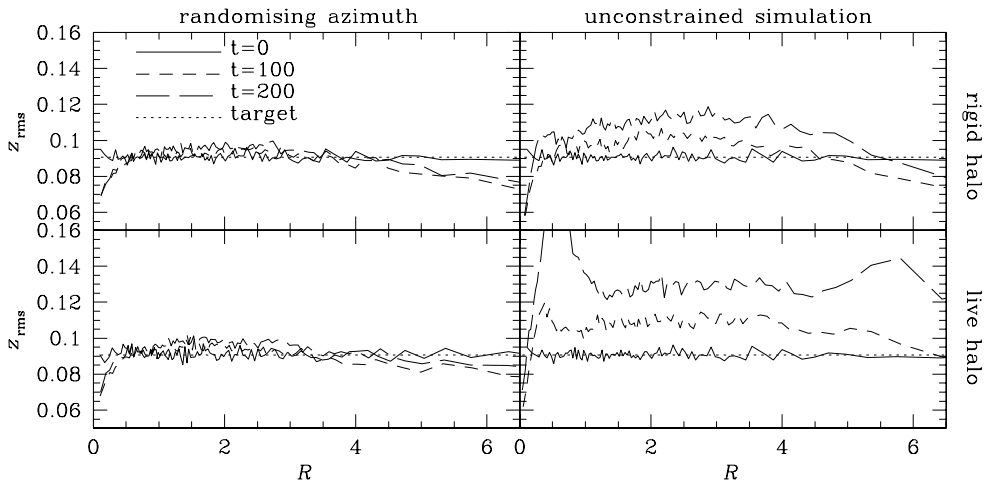}}}
  \caption{Evolution of the disk thickness in four different 
    simulations: constrained (by randomising the azimuth) and
    unconstrained, with rigid or live halo. In all cases, the
    initial velocity dispersion of the disc was set such that
    $Q=1.2$ everywhere. The lower-left panel corresponds to the
    middle panel of Fig.~\ref{fig:full:ld} (but note the different
    scales). \label{fig:disc:thick}
  }
\end{figure*}

In Fig.~\ref{fig:free:disc} we plot the surface density profiles of
discs with various initial $\sigma_R$ profiles. In each case $\sigma_R$
was defined such that $Q$ was constant across the disc.  Simulations are
shown with $Q=1.2$, 2, 3, and 4. For hotter discs (larger $Q$), it
becomes increasingly difficult to iterate the particle positions such
that the surface density is close to that desired, using the method
described in Section~\ref{sec:Method:disc}. This is why the initial
surface density of the $Q=4$ disc (and to a lesser extent the $Q=3$
disc) is not an exponential in radius. It should also be noted that the
motivation of warming an initially cold distribution function is
increasingly invalid as the disc becomes hotter. Moreover, such hot
axisymmetric discs are less realistic, given the observed fraction of
barred galaxies (which cannot form from such hot discs).

It has long been recognised that numerical simulations of cool discs are
much more susceptible to bar formation than those of warm discs
\citep[e.g.][]{Hohl1971}. This is what we observe in our models. The
scatter plot of particle positions for the $Q=1.2$ case,
Figure~\ref{fig:full:scat12}, shows spiral structure forming within 50
time units, and the formation of a strong bar within 100. This has the
effect of substantially altering the surface density profile,
dramatically increasing the disc's surface density in the inner
$\sim0.3$ scale lengths (as seen in Fig.~\ref{fig:free:disc}). The $Q=3$
disc shows no signs of developing spiral structure within 200 time units
(Fig.~\ref{fig:full:scat3}), and the density profile remains nearly
unchanged throughout the simulation.

\subsubsection{On the origin of disc thickening}
It is instructive to compare the disc thickening in the four different
cases we have looked at in this study (populated/rigid halo;
randomised azimuth/unconstrained model), see
Figure~\ref{fig:disc:thick}. Simulations which are constrained so that
non-axisymmetric modes are suppressed show hardly any disc thickening,
strongly suggesting that the dominant cause of disc thickening in our
models is the action of non-axisymmetric modes within the disc, rather
than the impact of halo particles crossing the disc.
\cite{Athanassoula2002} showed that a populated halo can stimulate bar
growth by absorbing angular momentum (which a rigid halo cannot). This
is a likely explanation for the greater increase in disc thickness
seen in the unconstrained simulation with live halo compared to that
with a rigid halo (right panels in Fig.~\ref{fig:disc:thick}).

Possibly the randomisation of the disk particles' azimuths is also a
direct cause of a reduction in disc thickening, because of he rapid
jumps it causes in the disc particles' positions (rather than just an
indirect cause, through preventing non-axisymmetric instabilities). Why
this would be the case, however, is unclear.

\section{Conclusions}\label{sec:iccon}
We have described and tested a new method for constructing an
equilibrium $N$-body representation of a galaxy with halo, disc and
bulge components.

We have used a distribution function for the halo and bulge based upon
\cite{Cuddeford1991} inversion, while the distribution function of the
disc is based on the work of \cite{Dehnen1999:DF}. One advantage over
previous methods is that our method avoids the use of a Maxwellian
approximation, and produces models which tend to stay very close to
their original states, though non-axisymmetric instabilities develop in
the disc if it is unconstrained and reasonably dynamically cold.

Another advantage of our method is that the density distributions of the
various components are straightforward to prescribe. That is, the way
the models are constructed guarantees that the equilibrium $N$-body
model has bulge, halo, and disc components the density profiles (as well
as the velocity-anisotropy profiles) of which follow the target models
very closely. Having said that, we have little control with our method
over the degree of non-sphericity of halo and bulge, introduced by the
disc's gravitational potential. We find that typically bulge and halo
are mildly oblate (axis ratio $\sim0.8$) in the region dominated by the
disc.

The choice of disc distribution function, while physically motivated,
and clearly in equilibrium (though potentially unstable to 
non-axisymmetric modes), causes problems
when creating a warm disc (Toomre's $Q\gtrsim4$) since it becomes
increasingly difficult to tailor the distribution function to the
desired density and velocity dispersion profiles. However, we are not
aware of any other method to create such warm $N$-body disc with a truly
exponential surface density profile.

We performed several tests to validate that the $N$-body model created
by our method meets our expectations. These tests suggest that the
growth of the disk thickness is predominantly driven by
non-axisymmetric modes in the disc itself rather than direct
interactions with halo particles.

The computer programs generated in the course of this project will be
publicly available under the \textsf{NEMO} computer package
(www.astro.umd.edu/nemo).

\section*{acknowledgement}
We are grateful for many useful discussions with Lia Athanassoula. PJM
acknowledges support by the UK Particle Physics and Astronomy Research
Council (PPARC) through a Research Student Fellowship and by an EU Marie
Curie Fellowship.  Astrophysics research at the University of Leicester
is also supported through a PPARC rolling grant.


\begin{thebibliography}{36}
\expandafter\ifx\csname natexlab\endcsname\relax\def\natexlab#1{#1}\fi

\bibitem[{{Athanassoula}(2002)}]{Athanassoula2002}
{Athanassoula} E., 2002, ApJL, 569, L83

\bibitem[{{Athanassoula}(2007)}]{Athanassoula2007}
{Athanassoula} E., 2007, MNRAS accepted, astro-ph/0703184

\bibitem[{{Athanassoula} \& {Misiriotis}(2002)}]{AM2002}
{Athanassoula} E., {Misiriotis} A., 2002, MNRAS, 330, 35

\bibitem[{{Barnes}(1988)}]{Barnes1988}
{Barnes} J.~E., 1988, ApJ, 331, 699

\bibitem[{{Boily} {et~al.}(2001){Boily}, {Kroupa}, \&
  {Pe{\~n}arrubia-Garrido}}]{BoilyKroupa2001}
{Boily} C.~M., {Kroupa} P., {Pe{\~n}arrubia-Garrido} J., 2001, NewA, 6, 27

\bibitem[{{Casertano} \& {Hut}(1985)}]{CasertanoHut1985}
{Casertano} S., {Hut} P., 1985, ApJ, 298, 80

\bibitem[{{Cuddeford}(1991)}]{Cuddeford1991}
{Cuddeford} P., 1991, MNRAS, 253, 414

\bibitem[{{Debattista} \& {Sellwood}(2000)}]{DebattistaSellwood2000}
{Debattista} V.~P., {Sellwood} J.~A., 2000, ApJ, 543, 704

\bibitem[{{Dehnen}(1999{\natexlab{a}})}]{Dehnen1999:Epi}
{Dehnen} W., 1999{\natexlab{a}}, AJ, 118, 1190

\bibitem[{{Dehnen}(1999{\natexlab{b}})}]{Dehnen1999:DF}
{Dehnen} W., 1999{\natexlab{b}}, AJ, 118, 1201

\bibitem[{{Dehnen}(2000)}]{Dehnen2000}
{Dehnen} W., 2000, ApJL, 536, L39

\bibitem[{{Dehnen}(2002)}]{Dehnen2002}
{Dehnen} W., 2002, Journal of Computational Physics, 179, 27

\bibitem[{{Dehnen} \& {Binney}(1998)}]{DehnenBinney1998}
{Dehnen} W., {Binney} J.~J., 1998, MNRAS, 294, 429

\bibitem[{{Eddington}(1916)}]{Eddington1916}
{Eddington} A.~S., 1916, MNRAS, 76, 572

\bibitem[{{Hernquist}(1990)}]{Hernquist1990}
{Hernquist} L., 1990, ApJ, 356, 359

\bibitem[{{Hernquist}(1993)}]{Hernquist1993}
{Hernquist} L., 1993, ApJS, 86, 389

\bibitem[{{Hernquist} \& {Ostriker}(1992)}]{HernquistOstriker1992}
{Hernquist} L., {Ostriker} J.~P., 1992, ApJ, 386, 375

\bibitem[{{Heyl} {et~al.}(1996){Heyl}, {Hernquist}, \&
  {Spergel}}]{Heyletal1996}
{Heyl} J.~S., {Hernquist} L., {Spergel} D.~N., 1996, ApJ, 463, 69

\bibitem[{{Hohl}(1971)}]{Hohl1971}
{Hohl} F., 1971, ApJ, 168, 343

\bibitem[{{Ideta} {et~al.}(2000){Ideta}, {Hozumi}, {Tsuchiya}, \&
  {Takizawa}}]{Idetaetal2000}
{Ideta} M., {Hozumi} S., {Tsuchiya} T., {Takizawa} M., 2000, MNRAS, 311, 733

\bibitem[{{Kazantzidis} {et~al.}(2004{\natexlab{a}}){Kazantzidis}, {Kravtsov},
  {Zentner}, {Allgood}, {Nagai}, \& {Moore}}]{Kazanzidisetal2004}
{Kazantzidis} S., {Kravtsov} A.~V., {Zentner} A.~R., {Allgood} B., {Nagai} D.,
  {Moore} B., 2004{\natexlab{a}}, ApJL, 611, L73

\bibitem[{{Kazantzidis} {et~al.}(2004{\natexlab{b}}){Kazantzidis}, {Magorrian},
  \& {Moore}}]{KazantzidisMagorrianMoore2004}
{Kazantzidis} S., {Magorrian} J., {Moore} B., 2004{\natexlab{b}}, ApJ, 601, 37

\bibitem[{{King}(1966)}]{King1966}
{King} I.~R., 1966, AJ, 71, 64

\bibitem[{{Klypin} {et~al.}(2002){Klypin}, {Zhao}, \&
  {Somerville}}]{KlypinZhaoSomerville2002}
{Klypin} A., {Zhao} H., {Somerville} R.~S., 2002, ApJ, 573, 597

\bibitem[{{Kuijken} \& {Dubinski}(1994)}]{KuijkenDubinski1994}
{Kuijken} K., {Dubinski} J., 1994, MNRAS, 269, 13

\bibitem[{{Kuijken} \& {Dubinski}(1995)}]{KuijkenDubinski1995}
{Kuijken} K., {Dubinski} J., 1995, MNRAS, 277, 1341

\bibitem[{{Merritt}(1985)}]{Merritt1985}
{Merritt} D., 1985, MNRAS, 214, 25P

\bibitem[{{Mihos} {et~al.}(1995){Mihos}, {Walker}, {Hernquist}, {Mendes de
  Oliveira}, \& {Bolte}}]{Mihosetal1995}
{Mihos} J.~C., {Walker} I.~R., {Hernquist} L., {Mendes de Oliveira} C., {Bolte}
  M., 1995, ApJL, 447, L87

\bibitem[{{Naab} {et~al.}(1999){Naab}, {Burkert}, \&
  {Hernquist}}]{Naabetal1999}
{Naab} T., {Burkert} A., {Hernquist} L., 1999, ApJL, 523, L133

\bibitem[{{Navarro} {et~al.}(1997){Navarro}, {Frenk}, \& {White}}]{NFW1997}
{Navarro} J.~F., {Frenk} C.~S., {White} S.~D.~M., 1997, ApJ, 490, 493

\bibitem[{{Osipkov}(1979)}]{Osipkov1979}
{Osipkov} L.~P., 1979, Pis ma Astronomicheskii Zhurnal, 5, 77

\bibitem[{{Sellwood}(1987)}]{Sellwood1987}
{Sellwood} J.~A., 1987, ARAA, 25, 151

\bibitem[{{Shu}(1969)}]{Shu1969}
{Shu} F.~H., 1969, ApJ, 158, 505

\bibitem[{{Spitzer}(1942)}]{Spitzer1942}
{Spitzer} L.~J., 1942, ApJ, 95, 329

\bibitem[{{Toomre}(1964)}]{Toomre1964}
{Toomre} A., 1964, ApJ, 139, 1217

\bibitem[{{Walker} {et~al.}(1996){Walker}, {Mihos}, \&
  {Hernquist}}]{Walkeretal1996}
{Walker} I.~R., {Mihos} J.~C., {Hernquist} L., 1996, ApJ, 460, 121

\bibitem[{{Widrow} \& {Dubinski}(2005)}]{WidrowDubinski2005}
{Widrow} L.~M., {Dubinski} J., 2005, ApJ, 631, 838

\end{thebibliography}

\label{lastpage}
\end{document}